\documentclass[sigconf]{acmart}

\AtBeginDocument{%
  \providecommand\BibTeX{{%
    \normalfont B\kern-0.5em{\scshape i\kern-0.25em b}\kern-0.8em\TeX}}}

\setcopyright{rightsretained}
\copyrightyear{2018}

\acmConference[CCBM '18]{\vspace{2pt}Poster at the 2nd Annual Center for Cellular and Biomolecular Machines Open House}{October 22, 2018}{University of California, Merced, USA}

\begin{document}

\title{Enabling Input on Tiny/Headless Systems Using Morse Code}

\author{Anna-Maria Gueorguieva}
\affiliation{%
  \institution{Human-Computer Interaction Group}
  \institution{University of California, Merced}
  \streetaddress{5200 N. Lake Road}
  \state{California}
  \city{Merced}
  \country{USA}}
  \postcode{95343}
\email{agueorguieva501267@muhsdstudents.org}

\author{Gulnar Rakhmetulla}
\affiliation{%
  \institution{Human-Computer Interaction Group}
  \institution{University of California, Merced}
  \streetaddress{5200 N. Lake Road}
  \state{California}
  \city{Merced}
  \country{USA}}
  \postcode{95343}
\email{grakhmetulla@ucmerced.edu}

\author{Ahmed Sabbir Arif}
\affiliation{%
  \institution{Human-Computer Interaction Group}
  \institution{University of California, Merced}
  \streetaddress{5200 N. Lake Road}
  \state{California}
  \city{Merced}
  \country{USA}}
  \postcode{95343}
\email{asarif@ucmerced.edu}

\renewcommand{\shortauthors}{Gueorguieva, Rakhmetulla, and Arif}

\begin{abstract}
 This paper presents results of a pilot study that explored the potential of Morse code as a method for text entry on mobile devices. In the study, participants without prior experience with Morse code reached 6.7 wpm with a Morse code keyboard in three short sessions. Learning was observed both in terms of text entry speed and accuracy, which suggests that the overall performance of the keyboard is likely to improve with practice.
\end{abstract}

\keywords{Morse code, accessibility, text entry, smartphone, virtual keyboard.}

\maketitle

\section{Introduction}
\begin{figure}
  \centering
  \includegraphics[width=0.6\linewidth]{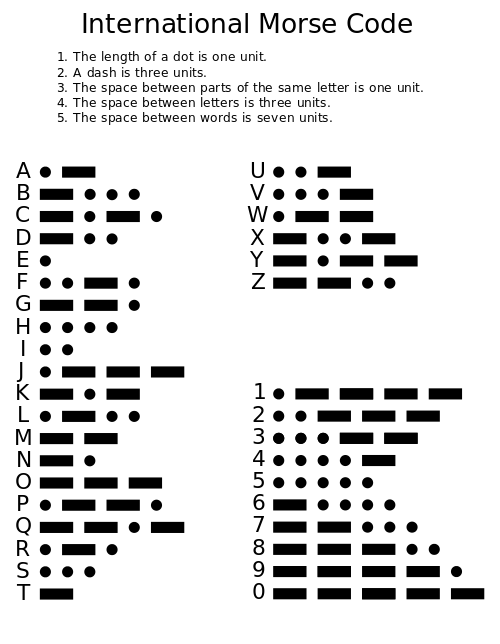}
  \caption{Morse code for the letters and numerals in the English language \cite{snodgrass_radio_1922}.}
  \label{fig:mcode}
\end{figure}

Entering text on devices with tiny or no displays (referred to as headless devices) remains a challenge. Conventional virtual keyboards are ineffective on devices with tiny displays, like smartwatches or smart thermostats, since the keys of these keyboards are too small for precise selection \cite{arif_survey_2016}. Entering text on a headless devices, such as a digital tangible in a tangible-tabletop system, is even more challenging due to the difficulties in verifying or correcting an input without visual feedback on a display. This work investigates the potential of Morse code as a method for text entry on these devices. Morse code encodes characters as standardized sequences of dots ($.$) and dashes ($-$) (Fig. \ref{fig:mcode}). Since it was originally designed for telegraphs, a device without a display \cite{snodgrass_radio_1922}, we hypothesize that it can enable users to enter text on both tiny and headless devices. In this work, however, we focus only on the learning of the code.

\section{Related Work}
Not much work has investigated the potential of Morse code as an input method on computer systems. The earliest work in this area studied the effectiveness of Morse code as an input method with a Z-80 based microcomputer \cite{levine_comparison_1986}. In the study, six participants yielded a 7.9 wpm entry speed after two months of training (2--3 sessions per week). A different work investigated if Morse code can be learned through passive haptic learning (PHL) using the bone conduction transducer on a Google Glass \cite{seim_tactile_2016}. In their study, participants reached an 8 wpm entry speed after four hours of exposure to passive stimuli. A follow up study demonstrated PHL of Morse code on a  smartwatch \cite{seim_smartwatch_2018}. Table \ref{tab:morse} summarizes the findings of these studies.

Several works have also explored the potential of Morse code with people with physical disabilities \cite{koester_text_2018} and with various channels, including eye-blink \cite{mukherjee_augmentative_2015}, foot stomping \cite{pedrosa_text_2014}, and tongue gestures \cite{sapaico_analysis_2011}. Some have used language models to improve text entry with Morse code  \cite{Kumiko_2005}. These, however, are outside the scope of this work.

\begin{table}[h]
  \centering
  \caption{Text entry speed (wpm) with Morse code reported in the literature. Here, ``PHL'' refers to passive haptic learning and ``$N$'' represents sample size.}
  \label{tab:morse}
  \begin{tabular}{lcccc}
    \toprule
    Reference&$N$&Session&Support&Speed\\
    \midrule
    \citet{levine_comparison_1986}&6&26--40&Cheat Sheet&7.9\\
    \citet{seim_tactile_2016}&12&4&PHL&8.0\\
  \bottomrule
\end{tabular}
\end{table}

\section{Morse Code Keyboard}
\begin{figure}
  \centering
  \includegraphics[width=0.8\linewidth]{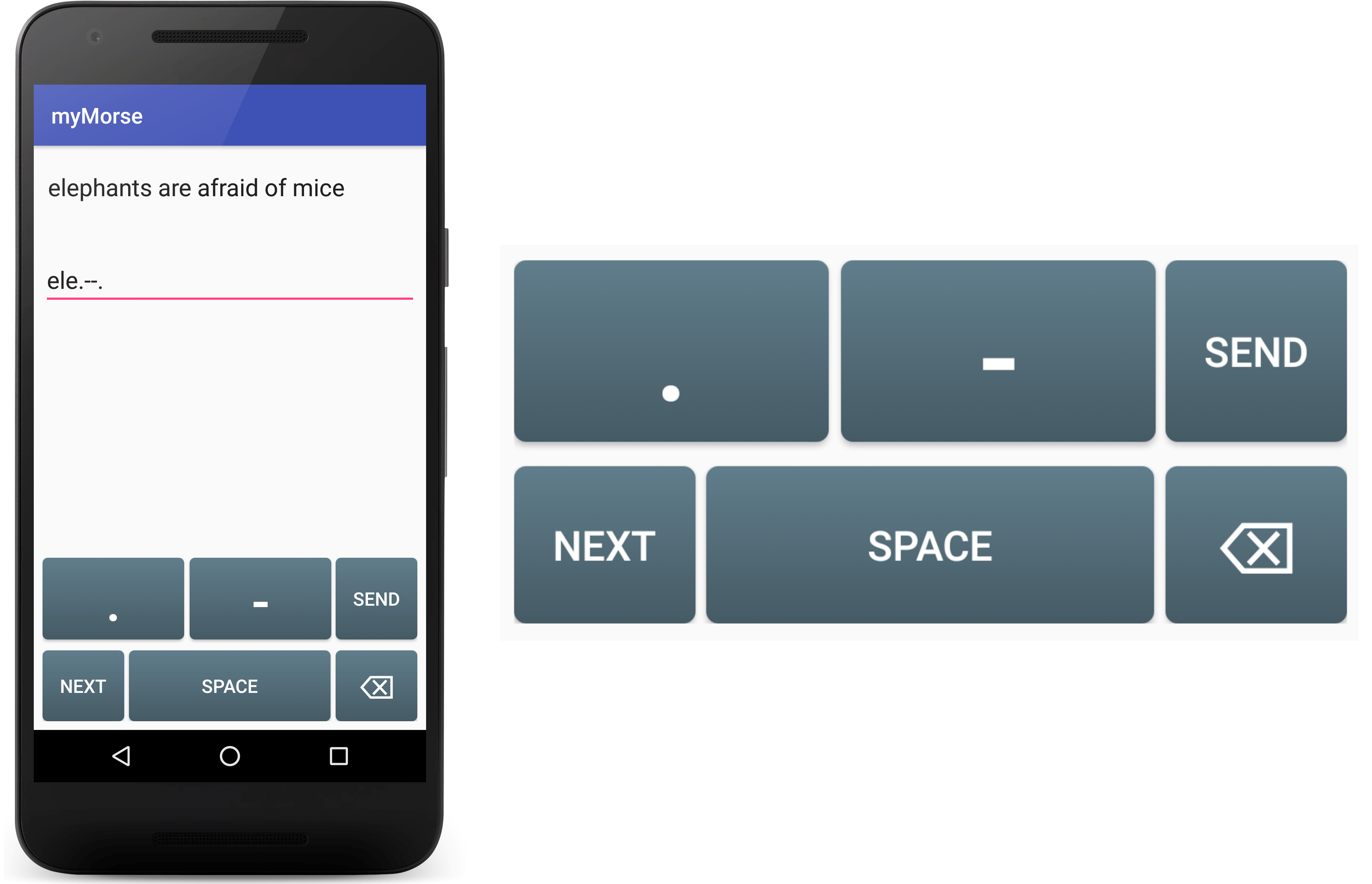}
  \caption{The device and keyboard used in the pilot study.}
  \label{fig:kb}
\end{figure}

We developed a simple virtual keyboard based on Morse code. It enables users to enter characters using sequences of dots ($.$) and dashes ($-$). The keyboard has dedicated keys for dot, dash, backspace, and space (Fig.~\ref{fig:kb}). To enter the letter ``R'', represented by ``$.-.$'' in Morse code, the user presses the respective keys in that exact sequence, followed by the SEND key, which terminates the input sequence. The user presses the NEXT key to terminate the current phrase and progress to the next one. The keyboard does not use a predictive system, thus does not auto-complete words, auto-correct incorrect words, or suggest the next probable words.

\section{Pilot Study}
We conducted a pilot study to investigate the learning of Morse code and the potential of Morse code as a method for text entry.

\subsection{Apparatus}
We used a Motorola Moto G$^5$ Plus smartphone (150.2$\times$74$\times$7.7 mm, 155 g) at 1080$\times$1920 pixels (Fig. \ref{fig:kb}). The Morse code keyboard was developed with the Android Studio 3.1, SDK 27. It automatically calculated all performance metrics and recorded all interactions with timestamps.

\subsection{Participants}
\begin{figure}[b]
  \centering
  \includegraphics[width=0.55\linewidth]{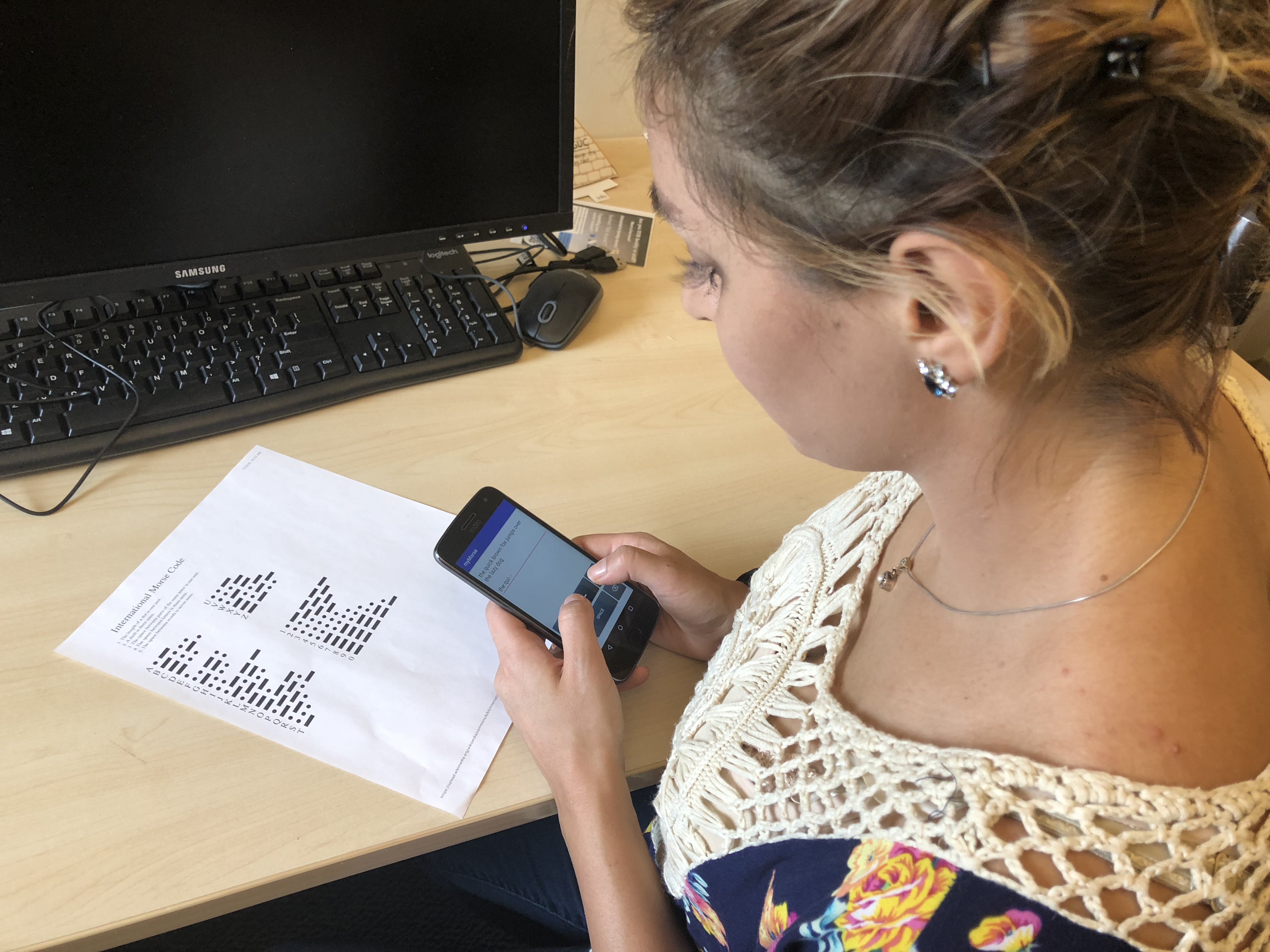}
  \caption{A volunteer transcribing text with the Morse code keyboard with the aid of a cheat sheet.}
  \label{fig:part}
\end{figure}

Two participants took part in the pilot study. Both of them were 25 years old. One of them was female and the other was male. They used both hands to hold the device and the thumbs to type (Fig. \ref{fig:part}). They were proficient in the English language. Both were experienced smartphone users (9 years' of experience) but had no prior experience with Morse code.

\subsection{Design}
We used a within-subjects design, where the independent variable was \textit{session} and the dependent variables were the commonly used \textit{words per minute} (wpm) and \textit{error rate} (ER) performance metrics \cite{arif_analysis_2009}. In summary, the design was as follows:

\begin{itemize}
  \item[] 2 participants $\times$
  \item[] 3 sessions (different days) $\times$
  \item[] 10 pangram entries = 60 entries, in total. 
\end{itemize}

\subsection{Procedure}
The study was conducted in a quiet room. On the first day, we explained the research to the participants, demonstrated Morse code and the custom keyboard, then enabled them to practice with the keyboard by entering free-form text for about five minutes. The first session started after that, where participants were instructed to enter the pangram ``the quick brown fox jumps over the lazy dog'' as fast and accurate as possible, then tap the NEXT key to submit the phrase. This was repeated for ten times. Participants could use a cheat sheet to look-up the code for a character (Fig. \ref{fig:part}). Error correction was disabled during the study. We instructed the participants to ignore all errors. Logging started from the first tap on the display and ended when participants pressed NEXT. The following sessions followed the same procedure, except for the demonstration and practice. The sessions were scheduled on consecutive days.

\section{Results}
We only report descriptive statistics due to the small sample size ($N = 2$) of the study.

\subsection{Entry Speed}
\begin{figure}[h]
  \centering
  \includegraphics[width=0.9\linewidth]{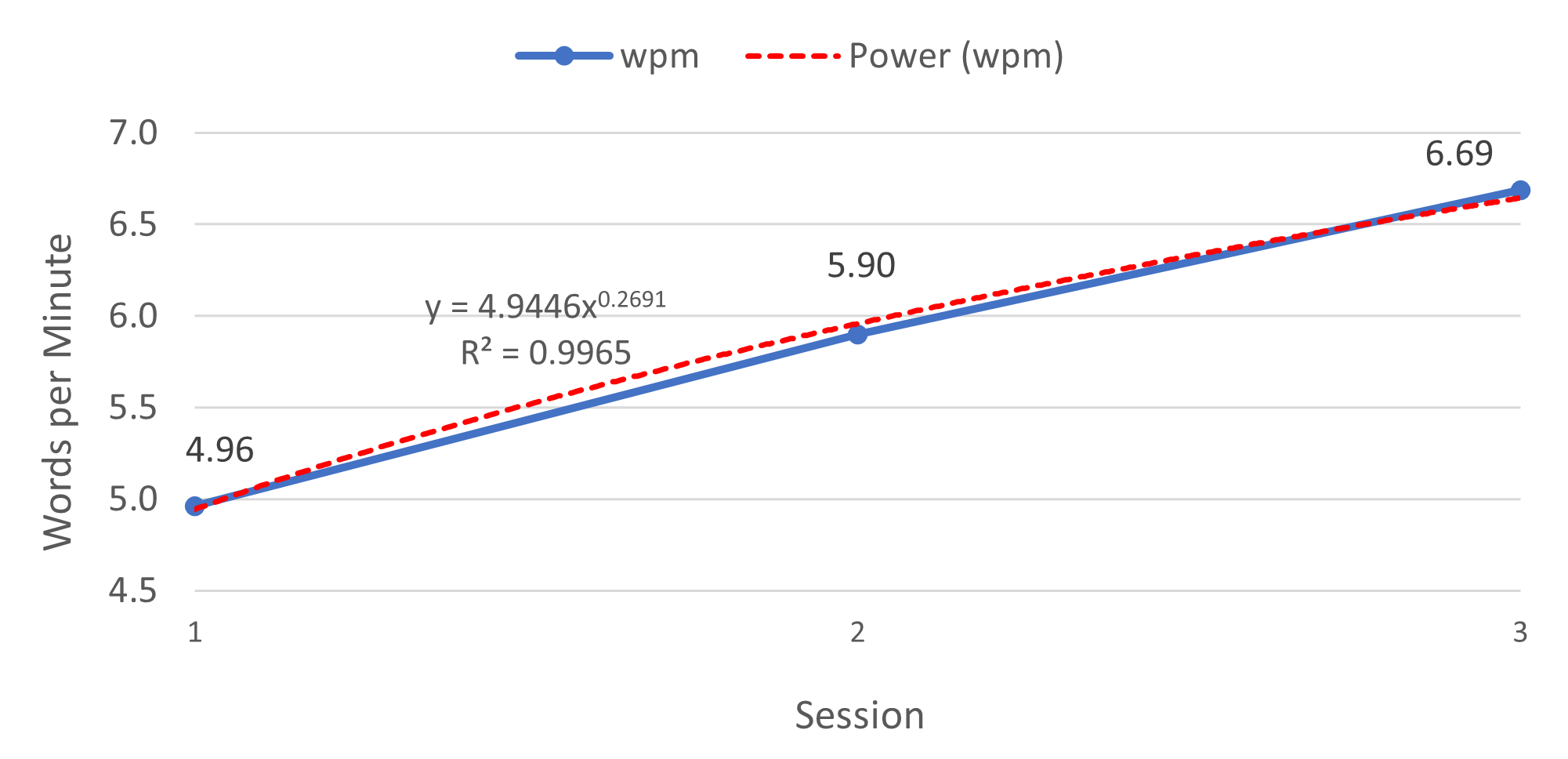}
  \caption{Average entry speed (wpm) in the three sessions fitted to a power trendline.}
  \label{fig:wpm}
\end{figure}

Participants yielded an average of 5.9 wpm (SD $= 1.1$). The average entry speed in the three sessions were 5.0 wpm (SD $= 0.6$), 5.9 wpm (SD $= 1.2$), and 6.7 wpm (SD $= 0.4$), respectively (Fig. \ref{fig:wpm}).

\subsection{Error Rate}
\begin{figure}[h]
  \centering
  \includegraphics[width=0.9\linewidth]{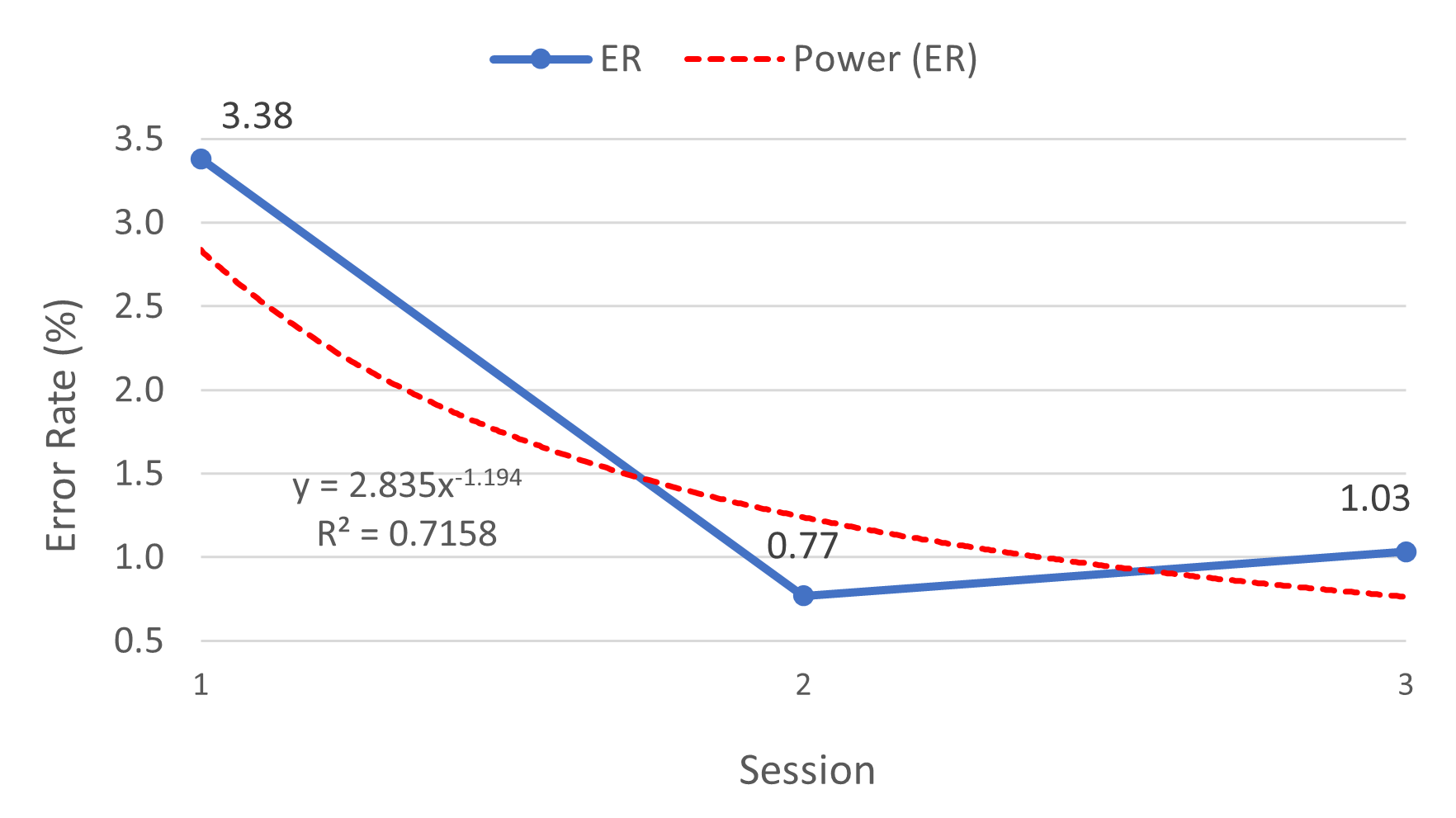}
  \caption{Average error rate (\%) in the three sessions fitted to a power trendline.}
  \label{fig:er}
\end{figure}

The average error rate in the study was 1.7\% (SD $= 2.6$). The average error rate in the three sessions were 3.4\% (SD $= 3.0$), 0.8\% (SD $= 1.2$), and 1.0\% (SD $= 2.5$), respectively (Fig. \ref{fig:er}).

\section{Discussion}
Participants yielded a competitive entry speed with the Morse code keyboard (Table \ref{tab:morse}), reaching an average of 6.7 wpm by the final session. It is inspiring that learning occurred even in such brief sessions of the study. Entry speed improved by 16\% in the second session and by 12\% in the third session compared to the preceding sessions. The average entry speed over session correlated well ($R^2 = 0.9965$) with the power law of practice \cite{snoddy_learning_1926}. This suggests that users are likely to get much faster with the method with practice. Participants made more errors in the first session (3.4\%), which reduced by 76\% in the second session. The error rate in the second and the final sessions were comparable (about 1\%). Innately, the average error rate over session moderately correlated ($R^2 = 0.7158$) with the power law of practice \cite{snoddy_learning_1926}. This suggests, unlike entry speed, error rate is unlikely to reduce substantially with practice. These findings indicate towards the possibility that Morse code could enable text entry on devices where using conventional input methods is impractical. Although it was not the focus of this work, Morse code could also enable people with various motor disabilities to enter text on computer systems.

\section{Limitations and Future Work}
There are several limitations of the work, which we will address in future studies. First, we evaluated the method on a smartphone where users could see their input. This visual feedback most likely improved their performance. It is unknown whether users will yield a similar performance with devices without a display. Second, we did not evaluate error correction in the study. Further investigation is needed to identify an effective method for identifying and correcting errors on devices without a display. Relevantly, studies showed that error correction is an integral part of text entry and difficulties in correcting errors significantly compromises users' experience with a method \cite{arif_predicting_2010}. Third, the custom keyboard includes dedicated keys for dots ($.$) and dashes ($-$), which may not be possible on all devices. Hence, investigation is needed to identify the best modes of interaction on devices that do not have physical or virtual keys. Some possibilities are using taps and long-taps, taps and hard-taps \cite{de_jong_one-press_2010,heo_forcetap:_2011,arif_use_2014,zeleznik_pop_2001}, or tap and flick \cite{baglioni_flick-and-brake:_2011} in place of dots and dashes. Finally, the study had a very small sample size ($N=2$). Replicating the study with a larger and more diverse sample will improve the generalizability of the findings.

\section{Conclusion}
We conducted a pilot study to investigate the potential of Morse code as a method for text entry on mobile devices. For this, we developed a simple Morse code keyboard. In the study, participants without prior experience with Morse code reached 6.7 wpm with the keyboard in three short sessions. Learning was observed both in terms of entry speed and accuracy, suggesting that the performance of the keyboard is likely to improve with practice.

\bibliographystyle{ACM-Reference-Format}
\bibliography{refs}

\end{document}